\documentstyle[aps]{revtex}
\date{}
\begin{document}
\title{Strong and electromagnetic decays of p-wave heavy baryons
$\Lambda_{c1}, \Lambda^*_{c1}$}
\author{Shi-Lin Zhu\\
Department of Physics, University of Connecticut, U-3046\\
2152 Hillside Road, Storrs,  CT 06269-3046
}
\maketitle
\begin{center}
\begin{minipage}{120mm}
\vskip 0.6in
\begin{center}{\bf Abstract}\end{center}
{We first calculate the binding energy, the pionic and electromagnetic couping constants
of the lowest lying p-wave heavy baryon doublet $\Lambda_{c1}$, $\Lambda_{c1}^*$
in the leading order of the heavy quark expansion. Then we calculate the
two-body decay widths with these couplings and compare our results with
other approaches. Our results are $\Gamma (\Lambda_{c1}\to \Sigma_c \pi,
\Sigma_c \gamma, \Sigma^*_c \gamma ) =2.7, 0.011, 0.001$ MeV
and $\Gamma (\Lambda^*_{c1}\to \Sigma_c \pi,
\Sigma_c \gamma, \Sigma^*_c \gamma, \Lambda_{c1} \gamma )
=33, 5, 6, 0.014$ keV
respectively. We find $\Lambda_{c1}, \Lambda^*_{c1}\to \Lambda_c \gamma$ is
strictly forbidden in the leading order of the heavy quark expansion.
At the order of ${\cal O} (1/m_c)$ their widths are
$36, 48$ keV respectively.

\vskip 0.5 true cm
PACS Indices: 14.20.Lq, 13.40.Hq, 13.75.Gx
}
\end{minipage}
\end{center}

\pagenumbering{arabic}
\section{ Introduction}
\label{sec1}

Now most of the ground state charm baryons have been found experimentally \cite{pdg}.
Important progress has been made in the search of orbitally excited heavy baryons.
The ARGUS \cite{argus}, E687 \cite{e687} and CLEO \cite{cleo5}) collaborations
have observed a pair of states in the channel $\Lambda_c^+\pi^+\pi^-$,
which were interpreted as the lowest lying orbitally excited states:
$\Lambda_{c1} (2593)$ with $J^P=\frac{1}{2}^-$ and $\Lambda_{c1}^{*}(2625)$ with
$J^P=\frac{3}{2}^-$. The total decay width of the $\Lambda_{c1} (2593)$ is
$3.6^{+2.0}_{-1.3}$ MeV while only an upper limit of $<1.9$ MeV has been set
for $\Lambda_c^*(2625)$ up to now \cite{pdg}.
Recently there emerges evidence for the $\Xi^{*+}_{c1}$ with $J^P=\frac{3}{2}^-$,
the strange partner of the $\Lambda_{c1}^* (2625)$. Its width is less than $2.4$ MeV.
In the near future much more data will be expected. We will focus on the strong and
electromagnetic decays of the $\Lambda_{c1}$ doublet since they are the only
well established states \cite{pdg}.

There exist many theoretical discussions on this topic.
In \cite{cho} the single pion and two pion strong decays and radiative decays of
the $\Lambda_{c1}$ doublet were discussed within the framework of heavy baryon chiral
perturbation theory. Due to unknown couplings constants in the chiral Lagrangian,
no actual decay widths were given. Within the same framework the pionic
decay widths were calculated assuming the heavy quark effective theory is still
valid for the strange quark \cite{dai}. The coupling constants in the chiral Lagrangian
were fixed using the p-wave strange baryon decay widths, which were later used
to predict the strong decays of the p-wave charm baryons \cite{dai}. The two pion
width of $\Lambda_{c1}$ was estimated to be around $2.5$ MeV, which was comparable
to the total one pion width $3.0$ MeV. And the decays of $\Lambda^*_{c1}$
was suppressed by more than an order \cite{dai}. In \cite{chow} the p-wave doublet
was treated as the bound state of the nucleon and heavy meson. It was found that the decays
$\Lambda_{c1}, \Lambda^*_{c1}\to \Lambda_c \gamma$ were suppressed due to the
kinematic suppression of the electric dipole moment \cite{chow}.
In \cite{yan} the constituent quark model was employed to study the orbitally excited
heavy baryons. Sum rules were derived to constrain the coupling constants.
The light front quark model, together with underlying $SU(2N_f)\times O(3)$ symmetry
for the light diquark system, was used to relate and analyse the pionic coupling
\cite{korner,t-2,t-4,t-1}. However, the results have strong dependence on the
constituent quark mass $m_q$. Varying $m_q$ from $220$ MeV to $340$ MeV, the
decay widths increase by more than a factor of two \cite{t-1}. Within the same
framework the electromagnetic decays of the p-wave baryons were calculated in \cite{t-3}.
In \cite{i-1} both strong and radiative decays were calculated using a
relativistic three-quark model. After this paper was submitted there appears an 
interesting paper discussing the radiative decays of the ground state heavy 
baryon multiplets in the framework of heavy baryon chiral perturbation theory. 
In some cases the loop corrections yield sizeable enhancement of the deca widths \cite{pich}.

It will be helpful to extract these pionic and photonic coupling constants
at the quark gluon level using QCD Lagrangian. We will treat this
problem using QCD sum rules (QSR) \cite{svz}, which are successful to extract
the low-lying hadron masses and couplings. In this approach
the nonperturbative effects are introduced via various condensates in the vacuum.
The light cone QCD sum rule (LCQSR) differs from the conventional short-distance QSR in that
it is based on the expansion over the twists of the operators. The main contribution
comes from the lowest twist operators. Matrix elements of nonlocal operators
sandwiched between a hadronic state and the vacuum define the hadron wave
functions. In the present case our sum rules involve with the pion and photon
wave function. When the LCQSR is used to calculate the coupling constant, the
double Borel transformation is always invoked so that the excited states and
the continuum contribution can be subtracted quite cleanly.
We have calculated the pionic and electromagnetic coupling constants and
decay widths of the ground state heavy hadrons \cite{zhu-meson,zhu-baryon,zhu-rad}
and possible hybrid heavy mesons \cite{zhu-hybrid}. In this work we extend the
same framework to study the strong and radiative decays of lowest p-wave
heavy baryons, i.e., $\Lambda_{c1}$ doublet.

Our paper is organized as follows: Section \ref{sec1} is an introduction.
In the next section we derive the mass sum rule. The light cone sum rules
for the pionic coupling constants are derived in Section \ref{sec3}.
Numerical analysis is presented. In Section \ref{sec4} we extend the
same framework to analyse the electromagnetic processes
$\Lambda_{c1}\to \Sigma_c\gamma$ etc. In Section \ref{sec5}
we discuss the processes $\Lambda_{c1}, \Lambda^*_{c1}\to \Lambda_c\gamma$ and
compare our results with other theoretical approaches.
The last section is a summary.
\section{The mass sum rules for the heavy hybrid mesons in HQET}
\label{sec2}

\subsection{Heavy quark effective theory}

The effective Lagrangian of the HQET, up to order $1/m_Q$, is
\begin{equation}
\label{Leff}
   {\cal L}_{\rm eff} = \bar h_v\,i v\!\cdot\!D\,h_v
   + \frac{1}{2 m_Q}\,{\cal K}
   + \frac{1}{2 m_Q}\,{\cal S}+{\cal O}(1/m_Q^2) \,,
\end{equation}
where $h_v(x)$ is the velocity-dependent field related to the
original heavy-quark field $Q(x)$ by
\begin{equation}
   h_v(x) = e^{i m_Q v\cdot x}\,\frac{1+\rlap/v}{2}\,Q(x)\;,
\end{equation}
$v_\mu$ is the heavy hadron velocity. ${\cal K}$ is the kinetic operator defined as
\begin{equation}
\label{kinetic}
{\cal K}=\bar h_v\,(i D_t)^2 h_v\;,
\end{equation}
where $D_t^\mu=D^\mu-(v\cdot D)\,v^\mu$, with
$D^\mu=\partial^\mu-i g\, A^\mu$ is the gauge-covariant derivative,
and $\cal S$ is the chromomagnetic operator
\begin{equation}
\label{pauli}
{\cal S}=\frac{g}{2}\,C_{mag}(m_Q/\mu)\;
   \bar h_v\,\sigma_{\mu\nu} G^{\mu\nu} h_v\;,
\end{equation}
where $C_{mag}=\displaystyle{\left(\alpha_s(m_Q)\over
\alpha_s(\mu)\right)^{3/{\beta}_0}}$, ${\beta}_0=11-2n_f/3$.
Note the heavy quark propogator has a simple form in coordinate space.
\begin{equation}\label{prop}
<0|T\{h_v (x), {\bar h}_v (0)\}|0>=\int_0^\infty dt \delta (x-vt){1+{\hat v}\over 2}\;.
\end{equation}

\subsection{The interpolating currents}

We introduce the interpolating currents for the relevant heavy baryons:
\begin{equation}
\label{lab1}
\eta_{\Lambda_c} (x) =\epsilon_{abc} [{u^a}^T(x) C\gamma_5 d^b (x)]  h_v^c (x) \, ,
\end{equation}
\begin{equation}
\label{si1}
\eta_{\Sigma_c^+} (x) =\epsilon_{abc} [{u^a}^T(x) C\gamma_{\mu} d^b (x)]
\gamma^{\mu}_t \gamma_5 h_v^c (x) \, ,
\end{equation}
\begin{equation}
\label{sig2}
\eta^{\mu}_{{\Sigma_c^{++}}^*} (x) =\epsilon_{abc} [{u^a}^T(x) C\gamma_{\nu} u^b (x)]
\Gamma_t^{\mu\nu} h_v^c (x) \, ,
\end{equation}
\begin{equation}
\label{la1}
\eta_{\Lambda_{c1}} (x) =\epsilon_{abc} [{u^a}^T(x) C\gamma_5 d^b (x)]
\gamma_t^\mu \gamma_5 D^t_\mu h_v^c (x) \, ,
\end{equation}
\begin{equation}
\label{la2}
\eta^\mu_{\Lambda_{c1}} (x) =\epsilon_{abc} [{u^a}^T(x) C\gamma_5 d^b (x)]
\Gamma_t^{\mu\nu} D^t_\nu h_v^c (x) \, ,
\end{equation}
where $a$, $b$, $c$ is the color index, $u(x)$, $d(x)$, $h_v(x)$ is the up,
down and heavy quark fields, $T$ denotes the transpose, $C$ is the charge conjugate
matrix, $\Gamma_t^{\mu\nu}=-g_t^{\mu\nu}+ {1\over 3} \gamma_t^{\mu}\gamma_t^{\nu}$,
$g_t^{\mu\nu}=g^{\mu\nu}-v^{\mu}v^{\nu}$, $\gamma_t^{\mu}=\gamma_{\mu}-
{\hat v}v^{\mu}$, and $v^{\mu}$ is the velocity of the heavy hadron.

The overlap amplititudes of the interpolating currents with the heavy baryons
are defined as:
\begin{equation}
\label{laap1}
\langle 0|\eta_{\Lambda_c} |\Lambda_c\rangle =f_{\Lambda_{c}}u_{\Lambda_c} \, ,
\end{equation}
\begin{equation}
\label{lap1}
\langle 0|\eta_{\Lambda_{c1}} |\Lambda_{c1}\rangle =f_{\Lambda_{c1}}u_{\Lambda_{c1}} \, ,
\end{equation}
\begin{equation}
\label{lap2}
\langle 0|\eta^\mu_{\Lambda^*_{c1}} |\Lambda^*_{c1}\rangle ={f_{\Lambda_{c1}^*}\over \sqrt{3}}
u^\mu_{\Lambda^*_{c1}} \, ,
\end{equation}
\begin{equation}
\label{lap3}
\langle 0|\eta_{\Sigma_c} |\Sigma_c\rangle =f_{\Sigma_c} u_{\Sigma_c}\, ,
\end{equation}
\begin{equation}
\label{lap42}
\langle 0|\eta^{\mu}_{\Sigma_c^*} |\Sigma_c^*\rangle ={f_{\Sigma_c^*}\over \sqrt{3}}
 u^{\mu}_{\Sigma_c^*} \, ,
\end{equation}
where $u^\mu_{\Lambda^*_{c1}}$, $u^{\mu}_{\Sigma_c^*}$ are the Rarita-Schwinger spinors in HQET.
In the leading order of HQET, $f_{\Sigma_c}= f_{\Sigma_c^*}$ and
$f_{\Lambda_{c1}}=f_{\Lambda_{c1}^*}$ due to heavy quark symmetry.

\subsection{The $\Lambda_{Q1}$ mass sum rules}

In order to extract the binding energy of the p-wave heavy baryons in the
leading order of HQET, we consider the correlators
\begin{equation}\label{co-1}
i\int d^4 x e^{ikx} \langle 0| T\{ \eta_{\Lambda_{c1}} (x),
{\bar \eta}_{\Lambda_{c1}} (0)\} |0\rangle
=\Pi (\omega) {1+{\hat v} \over 2}\;,
\end{equation}
with $\omega =k\cdot v$.

The dispersion relation for $\Pi (\omega )$ reads
\begin{equation}\label{dip-1}
\Pi ( \omega )=\int {\rho (s)\over s- \omega  -i\epsilon }ds\;,
\end{equation}
where $\rho (s)$ is the spectral density in the limit $m_Q \to \infty$.

At the phenomenological side
\begin{equation}
\Pi (\omega )= {f^2_{\Lambda_{c1}}
\over  \Lambda_{c1}-\omega}+\mbox{continuum} \;.
\end{equation}
In order to suppress the continuum and higher excited states
contribution we make Borel transformation with the variable $\omega$ to
(\ref{dip-1}). We have
\begin{equation}
\label{mass-1}
f_{\Lambda_{c1}}^2e^{-{\bar\Lambda_{\Lambda_{c1}}\over T}}
=\int_0^{s_0} \rho (s) e^{-{s\over T}}ds\;,
\end{equation}
where $\bar\Lambda_{\Lambda_{c1}}$ is the $\Lambda_{c1}$ binding energy of
in the leading order and $s_0$ is the continuum threshold.
Starting from $s_0$ we have modeled
the phenomenological spectral density with the parton-like one including
both the perturbative term and various condensates.

The spectral density $\rho (s)$ at the quark level reads,
\begin{equation}\label{spectral}
\rho (s) ={3\over 140\pi^4}s^7
-{1\over 384\pi^4} \langle g^2_sG^2\rangle s^3
+ {m_0^2 a^2\over 128\pi^4} \delta (s)
\end{equation}
where $a=-4\pi^2 \langle {\bar q} q \rangle =0.55$GeV$^3$,
$\langle g^2_sG^2\rangle =0.48$GeV$^4$,
$\langle {\bar q}g_s\sigma\cdot G q \rangle =m_0^2\langle {\bar q} q \rangle$,
and $m_0^2=0.8$ GeV$^2$.
An interesting feature of (\ref{spectral}) is that the gluon condensate is of
the opposite sign as the leading perturbative term, in contrast with the
ground state baryon mass sum rules. This may be interpreted as some kind of
gluon excitation since we are considering p-wave baryons. In the present case
the gluon in the covariant derivative also contributes to various condensates.

Two common approaches exist to extract the masses.
One is the derivative method.
\begin{equation}\label{derivative}
{\bar\Lambda}_{\Lambda_{c1}} ={\int_0^{s_0} s \rho (s) e^{-{s\over T}}ds \over
\int_0^{s_0} \rho (s) e^{-{s\over T}}ds } \; .
\end{equation}
The other is the fitting method, which involves with fitting the left hand side
and right hand side of Eq. (\ref{mass-1}) with the most suitable
parameters $\bar\Lambda_{\Lambda_{c1}}, f_{\Lambda_{c1}}, s^0$ in the working region
of the Borel parameter. With both methods we get consistent results,
\begin{eqnarray}
\label{para-1}
\bar\Lambda_{\Lambda_{c1}} &=&(1.1\pm 0.2)~\mbox{GeV}\;,\nonumber\\
f_{\Lambda_{c1}} &=&(0.025\pm 0.005)~\mbox{GeV}^4\;,\nonumber\\
s^0_{\Lambda_{c1}}&=&(1.45\pm 0.2)~\mbox{GeV}
\end{eqnarray}
in the working region $0.5-1.3$ GeV for the Borel parameter $T$.
For later use we also need the mass and overlapping amplitude of the
$\Sigma$, $\Lambda$ heavy baryon doublet, $\bar\Lambda_{\Sigma_c}$, $\bar\Lambda_{\Lambda_c}$,
$f_{\Sigma_c}$, $f_{\Lambda_c}$ in the leading order of $\alpha_s$ \cite{mass,dai-mass}.
\begin{eqnarray}
\label{para-2}
\bar\Lambda_{\Sigma_c}&=&(1.0\pm 0.1)~\mbox{GeV}\;,\nonumber\\
f_{\Sigma_c} &=&(0.04\pm 0.004)~\mbox{GeV}^3\;,\nonumber\\
s^0_{\Sigma_c}&=&(1.25\pm 0.15)~\mbox{GeV}
\end{eqnarray}
\begin{eqnarray}
\label{para-3}
\bar\Lambda_{\Lambda_c} &=&(0.8\pm 0.1)~\mbox{GeV}\;,\nonumber\\
f_{\Lambda_c} &=&(0.018\pm 0.004)~\mbox{GeV}^3\;,\nonumber\\
s^0_{\Lambda_c}&=&(1.2\pm 0.15)~\mbox{GeV}
\end{eqnarray}

\section{LCQSR for the pionic couplings}
\label{sec3}

\subsection{The correlator for pionic couplings}

We introduce the following amplitudes
\begin{equation}
\label{pi-1}
M(\Lambda_{c1}\to \Sigma_c \pi )=g_s {\bar u}_{\Sigma_c} u_{\Lambda_{c1}} \; ,
\end{equation}
\begin{equation}
\label{pi-2}
M(\Lambda^*_{c1}\to \Sigma_c \pi )=\sqrt{3}g_d
{\bar u}_{\Sigma_c}  \gamma_5 q^t_\mu {\hat q} u^\mu_{\Lambda^*_{c1}} \; ,
\end{equation}
\begin{equation}
\label{pi-3}
M(\Lambda_{c1}\to \Sigma^*_c \pi )=\sqrt{3} g^1_d {\bar u}^\mu_{\Sigma_c}
\gamma_5 q^t_\mu {\hat q} u_{\Lambda_{c1}} \; ,
\end{equation}
\begin{equation}
\label{pi-4}
M(\Lambda^*_{c1}\to \Sigma^*_c \pi )={\bar u}^\mu_{\Sigma_c}
[g'_s g^t_{\mu\nu} +3 g^2_d (q^t_\mu q^t_\nu -{1\over 3}g^t_{\mu\nu} q_t^2)]
u^\nu_{\Lambda^*_{c1}} \; ,
\end{equation}
where ${\hat q}=q_\mu \gamma^\mu$, $q_\mu$ is the pion momentum.
Only the first two decay processes are kinematically allowed.
Due to heavy quark symmetry, $g'_s =g_s$, $g^1_d =g^2_d =g_d$ in the
limit of $m_Q\to \infty$. In other words there are two independent
coupling constants correpsonding to s-wave and d-wave decays.
Note we are unable to determine the sign of $g_s$ and $g_d$. And
we are mainly interested in the decay widths of the p-wave heavy baryons.
In the following our convention is to let both couplings be positive.

We consider the following correlators
\begin{equation}
\label{pi-s-1}
 \int d^4x\;e^{ik\cdot x}\langle 0|T\left( \eta_{\Lambda_{c1}}(x)
{\bar \eta}_{\Sigma_c}(0) \right) |\pi(q)\rangle =
 {1+{\hat v}\over 2}  G_s (\omega,\omega')\;,
\end{equation}
\begin{equation}
\label{pi-s-d}
 \int d^4x\;e^{ik\cdot x}\langle 0|T\left(\eta^{\mu}_{\Lambda^*_{c1}}(x)
 {\bar \eta}_{\Sigma_c}(0)\right)|\pi(q)\rangle =
   {1+{\hat v}\over 2}q_\alpha q_\nu \Gamma_t^{\mu\alpha} \gamma^\nu_t \gamma_5
   G_d (\omega,\omega')\;,
\end{equation}
where $k^{\prime}=k-q$, $\omega=v\cdot k$, $\omega^{\prime}=v\cdot
k^{\prime}$ and $q^2=m_\pi^2=0$.

The function $G_{s,d}(\omega,\omega^{\prime})$ has the following pole terms
from double dispersion relation. For $G_s$ we have
\begin{eqnarray}
\label{pole}
{f_{\Lambda_{c1}}f_{\Sigma}g_s\over (\bar\Lambda_{\Lambda_{c1}}
-\omega')(\bar\Lambda_{\Sigma_c}-\omega)}+{c\over \bar\Lambda_{\Lambda_{c1}}
-\omega'}+{c'\over \bar\Lambda_{\Sigma_c}-\omega}\;.
\end{eqnarray}

\subsection{Pion light cone wave functions}

To go futher we need the two- and three-particle pion
light cone wave functions \cite{bely95}:
\begin{eqnarray}\label{phipi}
<\pi(q)| {\bar d} (x) \gamma_{\mu} \gamma_5 u(0) |0>&=&-i f_{\pi} q_{\mu}
\int_0^1 du \; e^{iuqx} (\varphi_{\pi}(u) +x^2 g_1(u) + {\cal O}(x^4) )
\nonumber \\
&+& f_\pi \big( x_\mu - {x^2 q_\mu \over q x} \big)
\int_0^1 du \; e^{iuqx}  g_2(u) \hskip 3 pt  , \label{ax} \\
<\pi(q)| {\bar d} (x) i \gamma_5 u(0) |0> &=& {f_{\pi} m_{\pi}^2 \over m_u+m_d}
\int_0^1 du \; e^{iuqx} \varphi_P(u)  \hskip 3 pt ,
 \label{pscal}  \\
<\pi(q)| {\bar d} (x) \sigma_{\mu \nu} \gamma_5 u(0) |0> &=&i(q_\mu x_\nu-q_\nu
x_\mu)  {f_{\pi} m_{\pi}^2 \over 6 (m_u+m_d)}
\int_0^1 du \; e^{iuqx} \varphi_\sigma(u)  \hskip 3 pt .
 \label{psigma}
\end{eqnarray}
\noindent

\begin{eqnarray}
& &<\pi(q)| {\bar d} (x) \sigma_{\alpha \beta} \gamma_5 g_s
G_{\mu \nu}(ux)u(0) |0>=
\nonumber \\ &&i f_{3 \pi}[(q_\mu q_\alpha g_{\nu \beta}-q_\nu q_\alpha g_{\mu \beta})
-(q_\mu q_\beta g_{\nu \alpha}-q_\nu q_\beta g_{\mu \alpha})]
\int {\cal D}\alpha_i \;
\varphi_{3 \pi} (\alpha_i) e^{iqx(\alpha_1+v \alpha_3)} \;\;\; ,
\label{p3pi}
\end{eqnarray}

\begin{eqnarray}
& &<\pi(q)| {\bar d} (x) \gamma_{\mu} \gamma_5 g_s
G_{\alpha \beta}(vx)u(0) |0>=
\nonumber \\
&&f_{\pi} \Big[ q_{\beta} \Big( g_{\alpha \mu}-{x_{\alpha}q_{\mu} \over q \cdot
x} \Big) -q_{\alpha} \Big( g_{\beta \mu}-{x_{\beta}q_{\mu} \over q \cdot x}
\Big) \Big] \int {\cal{D}} \alpha_i \varphi_{\bot}(\alpha_i)
e^{iqx(\alpha_1 +v \alpha_3)}\nonumber \\
&&+f_{\pi} {q_{\mu} \over q \cdot x } (q_{\alpha} x_{\beta}-q_{\beta}
x_{\alpha}) \int {\cal{D}} \alpha_i \varphi_{\|} (\alpha_i)
e^{iqx(\alpha_1 +v \alpha_3)} \hskip 3 pt  \label{gi}
\end{eqnarray}
\noindent and
\begin{eqnarray}
& &<\pi(q)| {\bar d} (x) \gamma_{\mu}  g_s \tilde G_{\alpha \beta}(vx)u(0) |0>=
\nonumber \\
&&i f_{\pi}
\Big[ q_{\beta} \Big( g_{\alpha \mu}-{x_{\alpha}q_{\mu} \over q \cdot
x} \Big) -q_{\alpha} \Big( g_{\beta \mu}-{x_{\beta}q_{\mu} \over q \cdot x}
\Big) \Big] \int {\cal{D}} \alpha_i \tilde \varphi_{\bot}(\alpha_i)
e^{iqx(\alpha_1 +v \alpha_3)}\nonumber \\
&&+i f_{\pi} {q_{\mu} \over q \cdot x } (q_{\alpha} x_{\beta}-q_{\beta}
x_{\alpha}) \int {\cal{D}} \alpha_i \tilde \varphi_{\|} (\alpha_i)
e^{iqx(\alpha_1 +v \alpha_3)} \hskip 3 pt . \label{git}
\end{eqnarray}
\noindent
The operator $\tilde G_{\alpha \beta}$  is the dual of $G_{\alpha \beta}$:
$\tilde G_{\alpha \beta}= {1\over 2} \epsilon_{\alpha \beta \delta \rho}
G^{\delta \rho} $; ${\cal{D}} \alpha_i$ is defined as
${\cal{D}} \alpha_i =d \alpha_1
d \alpha_2 d \alpha_3 \delta(1-\alpha_1 -\alpha_2
-\alpha_3)$.
Due to the choice of the
gauge  $x^\mu A_\mu(x) =0$, the path-ordered gauge factor
$P \exp\big(i g_s \int_0^1 du x^\mu A_\mu(u x) \big)$ has been omitted.

The wave function $\varphi_{\pi}(u)$ is associated with the leading twist 2
operator, $g_1(u)$ and $g_2(u)$ correspond to twist 4 operators, and $\varphi_P(u)$ and
$\varphi_\sigma (u)$ to twist 3 ones.
The function $\varphi_{3 \pi}$ is of twist three, while all the wave
functions appearing in eqs.(\ref{gi}), (\ref{git}) are of twist four.
The wave functions $\varphi (x_i,\mu)$ ($\mu$ is the renormalization point)
describe the distribution in longitudinal momenta inside the pion, the
parameters $x_i$ ($\sum_i x_i=1$)
representing the fractions of the longitudinal momentum carried
by the quark, the antiquark and gluon.

The wave function normalizations immediately follow from the definitions
(\ref{phipi})-(\ref{git}):
$\int_0^1 du \; \varphi_\pi(u)=\int_0^1 du \; \varphi_\sigma(u)=1$,
$\int_0^1 du \; g_1(u)={\delta^2/12}$,
$\int {\cal D} \alpha_i \varphi_\bot(\alpha_i)=
\int {\cal D} \alpha_i \varphi_{\|}(\alpha_i)=0$,
$\int {\cal D} \alpha_i \tilde \varphi_\bot(\alpha_i)=-
\int {\cal D} \alpha_i \tilde \varphi_{\|}(\alpha_i)={\delta^2/3}$,
with the parameter $\delta$ defined by
the matrix element:
$<\pi(q)| {\bar d} g_s \tilde G_{\alpha \mu} \gamma^\alpha u |0>=
i \delta^2 f_\pi q_\mu$.

\subsection{The pionic sum rules}

Now the expressions of $G_s$, $G_d$ at the quark level read,
\begin{eqnarray}\label{quark1}\nonumber
G_s( \omega, \omega')= i f_\pi \int_0^{\infty} dt
\int_0^1 du e^{i (1-u) \omega t } e^{i u \omega' t }
\{
{6\mu_\pi\over \pi^2 t^4} \varphi_P (u)
+{\mu_\pi\over 3\pi^2 t^4} \left( 3\varphi'_\sigma (u)
+[u \varphi_\sigma (u)]^{''} \right) &\\  \nonumber
+(\langle {\bar q} q \rangle +{t^2\over 16}
\langle {\bar q}g_s\sigma\cdot G q \rangle ) g_2(u)
+(\langle {\bar q} q \rangle +{t^2\over 16}
\langle {\bar q}g_s\sigma\cdot G q \rangle )
{ [u \varphi_\pi (u)]^{''}+ t^2 [uG_2(u) +u g_1(u)]^{''} \over 3t^2}
 \} &\\ \nonumber
+{i\over \pi^2}f_{3\pi} \int {dt \over t^2} \int_0^1du (1-u)
\int {\cal D} \alpha_i
e^{i \omega t [1-(\alpha_1 +u\alpha_3)] }
e^{i \omega' t (\alpha_1 +u\alpha_3)}
[(q\cdot v)^2 -it (q\cdot v)^3 (\alpha_1+u \alpha_3)]
\varphi_{3\pi} (\alpha_i) &\\
-{2i\over \pi^2}f_{3\pi} \int {dt \over t^2} \int_0^1du u
\int {\cal D} \alpha_i
e^{i \omega t [1-(\alpha_1 +u\alpha_3)]}
e^{i \omega' t (\alpha_1 +u\alpha_3)}
(q\cdot v)^2 \varphi_{3\pi} (\alpha_i) &  \;,
\end{eqnarray}
\begin{eqnarray}\label{quark2}\nonumber
G_d( \omega, \omega')= i f_\pi \int_0^{\infty} dt
\int_0^1 du u e^{i (1-u) \omega t } e^{i u \omega' t }
\{
{\mu_\pi\over 3\pi^2 t^2} \varphi_\sigma (u) &\\   \nonumber
+{1\over 3}(\langle {\bar q} q \rangle +{t^2\over 16}
\langle {\bar q}g_s\sigma\cdot G q \rangle )
\left(  \varphi_\pi (u)+ t^2 [G_2(u) + g_1(u)] \right)
\} &\\ \nonumber
-{i\over \pi^2}f_{3\pi} \int {dt \over t^2} \int_0^1du (1-u)
\int {\cal D} \alpha_i
e^{i \omega t [1-(\alpha_1 +u\alpha_3)]}
e^{i \omega' t (\alpha_1 +u\alpha_3)}
[1 -it (q\cdot v) (\alpha_1+u \alpha_3)]
\varphi_{3\pi} (\alpha_i) &\\
-{i\over \pi^2}f_{3\pi} \int {dt \over t^2} \int_0^1du u
\int {\cal D} \alpha_i
e^{i \omega t [1-(\alpha_1 +u\alpha_3)]}
e^{i \omega' t (\alpha_1 +u\alpha_3)}
\varphi_{3\pi} (\alpha_i) &  \;,
\end{eqnarray}
where $\mu_{\pi}=1.65$GeV, $f_{\pi}=132$MeV, $F'(u)={dF(u)\over du}$
and $F''(u)={d^2F(u)\over du^2}$. There are two three particle terms
in the form of $\varphi_{3\pi}$ in (\ref{quark1}), (\ref{quark2}).
The gluon arises from the light quark propagator in the first term
and from the covariant derivative in the second term.
For large euclidean values of $\omega$ and $\omega'$
this integral is dominated by the region of small $t$, therefore it can be
approximated by the first few terms with lowest twists.

After Wick rotations and making double Borel transformation
with the variables $\omega$ and $\omega'$
the single-pole terms in (\ref{pole}) are eliminated.
Subtracting the continuum contribution which is modeled by the
dispersion integral in the region
$s ,s' \ge s_0$, we arrive at:
\begin{eqnarray}\label{g-s}
\nonumber
 g_s f_{\Lambda_{c1}} f_{\Sigma_c} =& {f_{\pi}\over \pi^2}
 e^{ {\bar\Lambda_{\Lambda_{c1}} +\bar\Lambda_{\Sigma_c}  \over 2T }}
 \{
6\mu_{\pi}\varphi_P(u_0)T^5 f_4({s_0\over T})
+{f_{3\pi}\over f_\pi} (2I_3-I_4-I_6)T^5 f_4({s_0\over T})\\ \nonumber
&+{\mu_\pi\over 3} \left( 3\varphi'_\sigma (u_0)
+[u \varphi_\sigma (u)]_{u_0}^{''} \right) T^5 f_4({s_0\over T})
+{a\over 12}[u\varphi_\pi (u)]^{''}_{u_0} (1-{m_0^2\over 16T^2})T^3 f_2({s_0\over T})\\
&-{a\over 4}\left( g_2 (u_0) +{1\over 3} [uG_2(u) +u g_1(u)]_{u_0}^{''} \right)
(1-{m_0^2\over 16T^2})T f_0({s_0\over T}) \} \;,
\end{eqnarray}
where $f_n(x)=1-e^{-x}\sum\limits_{k=0}^{n}{x^k\over k!}$ is the factor used
to subtract the continuum, $s_0$ is the continuum threshold.
$u_0={T_1 \over T_1 + T_2}$,
$T\equiv {T_1T_2\over T_1+T_2}$, $T_1$, $T_2$ are the Borel parameters.
The functions $I_i$ are defined below.
In obtaining (\ref{g-s}) we have used the Borel transformation formula:
${\hat {\cal B}}^T_{\omega} e^{\alpha \omega}=\delta (\alpha -{1\over T})$
and integration by parts to absorb the factors $(q\cdot v)$ and $1/(q\cdot v)$.
In this way we arrive at the simple form after double Borel transformation.

Similarly we have:
\begin{eqnarray}\label{g-d}\nonumber
 g_d f_{\Lambda^*_{c1}} f_{\Sigma_c} =& {f_{\pi}\over \pi^2}
 e^{ {\bar\Lambda_{\Lambda^*_1} +\bar\Lambda_{\Sigma_c}  \over 2T }}
 \{
{\mu_\pi\over 3} u_0 \varphi_\sigma (u_0)T^3 f_2({s_0\over T})
-{f_{3\pi}\over f_\pi} (I_1+I_2+I_5)T^3 f_2({s_0\over T})\\
&+{a\over 12}u_0\varphi_\pi (u_0) (1-{m_0^2\over 16T^2})T f_0({s_0\over T})
-{a\over 12T} u_0[G_2(u_0) + g_1(u_0)](1-{m_0^2\over 16T^2})\}
\;.
\end{eqnarray}

The functions $G_2(u_0)$, $I_i$ are defined as:
\begin{equation}
G_2 (u_0)=-\int_0^{u_0} g_2(u)du \; ,
\end{equation}
\begin{equation}
I_1 =\int_0^{u_0} d\alpha_1 \int_0^{1-u_0} d\alpha_2
{u_0-\alpha_1\over \alpha^2_3} \varphi_{3\pi} (\alpha_i)
 \; ,
\end{equation}
\begin{equation}
I_2 =\int_0^{u_0} d\alpha_1 \int_0^{1-u_0} d\alpha_2
{1-u_0-\alpha_2\over \alpha^2_3} \varphi_{3\pi} (\alpha_i)
 \; ,
\end{equation}
\begin{eqnarray}\nonumber
I_3 =\int_0^{u_0} d\alpha_1
{d\over d\alpha_3}[{ \varphi_{3\pi} (\alpha_1, 1-\alpha_1-\alpha_3, \alpha_3)
\over \alpha_3}]|_{\alpha_3=u_0-\alpha_1} &\\
-\int_0^{u_0} d\alpha_1 {\varphi_{3\pi} (\alpha_1, 1-u_0, u_0-\alpha_1)
\over (u_0 -\alpha_1)^2 }
+\int_0^{1-u_0} d\alpha_2 {\varphi_{3\pi} (u_0,\alpha_2, 1-u_0-\alpha_2)
\over (1-u_0 -\alpha_2)^2 } &
 \; ,
\end{eqnarray}
\begin{eqnarray}\nonumber
I_4 =\int_0^{1-u_0} {d\alpha_3 \over \alpha_3}
[{ d\varphi_{3\pi} (\alpha_1, 1-\alpha_1-\alpha_3, \alpha_3)
\over d\alpha_1}]|_{\alpha_1=u_0} &\\
+\int_0^{u_0} d\alpha_1 {\varphi_{3\pi} (\alpha_1, 1-u_0, u_0-\alpha_1)
\over (u_0 -\alpha_1)^2 }
-\int_0^{1-u_0} d\alpha_2 {\varphi_{3\pi} (u_0,\alpha_2, 1-u_0-\alpha_2)
\over (1-u_0 -\alpha_2)^2 }&
 \; ,
\end{eqnarray}
\begin{equation}
I_5 =-\int_0^{1-u_0} u_0 {d\alpha_3\over \alpha_3}
\varphi_{3\pi} (u_0,1-u_0-\alpha_3,\alpha_3 )
+\int_0^{u_0} d\alpha_1 \int_0^{1-u_0} d\alpha_2
{2u_0-1+\alpha_2\over \alpha^2_3} \varphi_{3\pi} (\alpha_i)
 \; ,
\end{equation}
\begin{eqnarray}\nonumber
I_6 = { d[\alpha_1 \varphi_{3\pi} (\alpha_1, 1-\alpha_1-\alpha_3, \alpha_3)]
\over d \alpha_1}|_{\alpha_3=1-u_0}^{\alpha_1=u_0}&\\ \nonumber
-\int_0^{1-u_0} d\alpha_3
{d^2\over d{\alpha_1}^2}[ \varphi_{3\pi} (\alpha_1, 1-\alpha_1-\alpha_3, \alpha_3)
{\alpha_1\over \alpha_3}]|_{\alpha_1=u_0} &\\ \nonumber
+[ \varphi_{3\pi} (\alpha_1, 1-\alpha_1-\alpha_3, \alpha_3)
{\alpha_3-\alpha_1\over \alpha^2_3}]|^{\alpha_1=0}_{\alpha_3=u_0} &\\ \nonumber
+\int_0^{u_0} d\alpha_3 \int_0^{u_0-\alpha_3} d\alpha_1
{d\over d\alpha_1}[ \varphi_{3\pi} (\alpha_1, 1-\alpha_1-\alpha_3, \alpha_3)
{\alpha_3-\alpha_1\over \alpha^2_3}] &\\ \nonumber
+[ \varphi_{3\pi} (\alpha_1, 1-\alpha_1-\alpha_3, \alpha_3)
{\alpha_3-\alpha_1\over \alpha^2_3}]|^{\alpha_1=u_0}_{\alpha_3=1-u_0} &\\ \nonumber
-\int_0^{1-u_0} d\alpha_3
{d\over d\alpha_1}[ \varphi_{3\pi} (\alpha_1, 1-\alpha_1-\alpha_3, \alpha_3)
{\alpha_3-\alpha_1\over \alpha^2_3}]|_{\alpha_1=u_0} & \\ \nonumber
-2[{\varphi_{3\pi} (\alpha_i) \over \alpha_3}]|^{\alpha_1=0}_{\alpha_3=u_0}
-2\int_0^{u_0} d\alpha_3 \int_0^{u_0-\alpha_3} d\alpha_1
{d\over d\alpha_1}[ {\varphi_{3\pi} (\alpha_1, 1-\alpha_1-\alpha_3, \alpha_3)
\over \alpha_3}] & \\ \nonumber
-2\int_0^{u_0} d\alpha_3\int_0^{u_0-\alpha_3} d\alpha_1
{\varphi_{3\pi} (\alpha_1, 1-\alpha_1-\alpha_3, \alpha_3) \over \alpha_3^2 }& \\
+2\int_0^{1-u_0} d\alpha_3 \int_0^{1-u_0-\alpha_3} d\alpha_2
{\varphi_{3\pi} (1-\alpha_2-\alpha_3, \alpha_2, \alpha_3)
\over \alpha_3^2 } &
\; ,
\end{eqnarray}
where $\alpha_3, \alpha_1$ are the longitudinal momentum fraction of gluon and 
down quark inside the pion respectively.

\subsection{Determination of the parameters for pionic sum rules}

The mass difference between $\Lambda_{c1}$ and $\Sigma_c$ is only about $0.1$GeV
in the leading order of HQET. And the values of the Borel parameter $T_1, T_2$
is around $2$ GeV in the working region. So we choose to work at the symmetric
point $T_1=T_2=2T$, i.e., $u_0 ={1\over 2}$, which diminishes the uncertainty
arising from the pion wave functions and enables a rather clean subtraction
of the continuum contribution.

The pion wave functions and their values at the middle point
are discussed in \cite{bely95,ball-pi,zhu-pi}.
At the scale $\mu =1.0$GeV the values of the various functions appearing
in (\ref{g-s})-(\ref{g-d}) at $u_0={1\over2}$ are:
$\varphi_\pi(u_0)=(1.5\pm 0.2)$ \cite{zhu-pi},
$\varphi_P(u_0)=1.142$, $\varphi_\sigma(u_0)=1.463$, $g_1(u_0)=0.034 $GeV$^2$,
$G_2(u_0)=0.02 $GeV$^2$ \cite{ball-pi}, $\varphi'_\sigma(u_0)=0$, $g_2 (u_0)=0$,
$[u\varphi_\pi (u)]^{''}_{u=u_0}=[u\varphi_\sigma (u)]^{''}_{u=u_0}=-6$,
$[ug_1(u)+uG_2(u)]^{''}_{u=u_0}=-0.29$,
$I_1=1.17$, $I_2=1.17$, $I_3=31.9$, $I_4=-31.9$, $I_5=-1.64$, $I_6= 247.5$,
$f_{3\pi}=0.0035$GeV$^2$.
We have used the forms in \cite{ball-pi} for $\varphi_{3\pi}(\alpha_i)$
to calculate integrals $I_i $.
The three particle wave functions are known to next order in the conformal
spin expansion up to now. The second derivatives need knowledge of the detailed
shape of the pion wave functions at the middle point. Various sources indicate
$\varphi_\pi (u)$ is very close to the asymptotic form \cite{zhu-pi}, which
is exactly known. Based on these considerations we have employed the
asymptotic forms to extract the second derivatives for $\varphi_\sigma (u)$ and
$\varphi_\pi (u)$.

\subsection{Numerical analysis of pionic sum rules}

Note the spectral density of the sum rule (\ref{g-s})-(\ref{g-d})
is either proptional to $s^2$ or $s^4$, the continuum has to be
subtracted carefully. We use $s_0=(1.3\pm 0.15)$ GeV, which is the
average of the thresholds of the $\Lambda_{c1}$ and $\Sigma_c$
mass sum rules. The variation of $g_{s,d}$ with the Borel parameter $T$ and $s_0$ is
presented in Fig. 1 and Fig. 2. The curves correspond to
$s_0 =1.2, 1.3, 1.4$GeV from bottom to top respectively.
Stability develops for these sum rules
in the region $0.5$ GeV $<$$T$$<$$1.5$ GeV, we get:
\begin{eqnarray}
\label{result}
 &&g_s f_{\Lambda_{c1}} f_{\Sigma} =(0.5\pm 0.3)\times 10^{-3}\mbox{GeV}^7\;,\\
 &&g_d f_{\Lambda^*_1} f_{\Sigma}=(2.8\pm 0.6)\times 10^{-3}\mbox{GeV}^5\;,
\end{eqnarray}
where the errors refers to the variations with $T$ and
$s_0$ in this region. And the central value corresponds
to $T=1$GeV and $s_0 =1.3$GeV.

Combining (\ref{para-1}), (\ref{para-2}) we get
\begin{eqnarray}
\label{final}
 &&g_s  =(0.5\pm 0.3)\;,\\
 &&g_d =(2.8\pm 0.6)\mbox{GeV}^{-2}\;.
\end{eqnarray}
We collect the values of the pionic couplings from various approaches
TABLE I. Note in our notation $3g_d$ corresponds to those in \cite{i-1}.

We use the following formulas to calculate the pionic decay widths of p-wave heavy baryons.
\begin{equation}
\Gamma (\Lambda_{c1}\to \Sigma_c \pi ) = {g_s^2\over 2\pi}
{m_{\Sigma_c}\over m_{\Lambda_{c1}}} |q| \; ,
\end{equation}
\begin{equation}
\Gamma (\Lambda^*_{c1}\to \Sigma_c \pi ) = {g_d^2\over 2\pi}
{m_{\Sigma_c}\over m_{\Lambda^*_{c1}}} |q|^5 \; ,
\end{equation}
where $|q|$ is the pion decay momentum. We use the values
$m_{\Lambda_{c1}}=2.593$ GeV, $m_{\Lambda^*_{c1}}=2.625$ GeV,
$m_{\Sigma_c}=2.452$ GeV \cite{pdg}. In the $\Lambda_{c1}$ decays
due to isospin symmetry violations of the pion and $\Sigma_c$ multiplet
masses, the pion decay momentum is $17, 23, 32$ MeV for the final states
$\Sigma_c^{++} \pi^-, \Sigma_c^{0} \pi^+, \Sigma_c^{+} \pi^0$ respectively.
This effect causes significant difference in the decay widths,
which are collected in TABLE II. Summing all the three isospin channels
we get $\Gamma (\Lambda_{c1}\to \Sigma_c \pi )=2.7$ MeV and
$\Gamma (\Lambda^*_{c1}\to \Sigma_c \pi )=33$ keV. The later is nearly suppressed by two
oders of magnitude due to d-wave decays.

>From TABLE II we see that our results are numerically close to those from fixing
the unknown coupling constants from the p-wave strange baryon strong decay widths
assuming heavy quark
effective theory could be extended to the strange quark case \cite{dai}. The values
of d-wave decay widths from the above approach and ours are much smaller than
those from the quark models \cite{i-1,t-2,t-1}. As for the s-wave decays various
approaches yield consistent results.

\section{Radiative decays of p-wave heavy baryons}
\label{sec4}

\subsection{The correlator}
The light cone photon wave functions have been used to discuss
radiative decay processes in \cite{mendel,aliev,ali,stoll,ball,zhu-rad}
in the framework of QCD sum rules. We extend the same formalism to
extract the electromagnetic coupling consants for the $\Lambda_{Q1}$
doublet decays.

The radiative coupling constants are defined through the following amplitudes:
\begin{equation}
\label{coup-1}
M(\Lambda_{c1}\to\Sigma_c\gamma )=
e\epsilon_{\beta \nu \rho \mu} q^\beta {e^\nu}^*
{\bar u}_{\Sigma_c}
[f_s g_t^{\rho\alpha} +f_d q^\alpha v^\rho]
\gamma_\alpha^t \gamma^\mu_t
u_{\Lambda_{c1}} \; ,
\end{equation}
\begin{equation}
\label{coup-2}
M(\Lambda_{c1}\to\Sigma^*_c\gamma )=
\sqrt{3}e\epsilon_{\beta \nu \rho \mu} q^\beta {e^\nu}^*
{\bar u}^\alpha_{\Sigma^*_c}
[f^1_s g_t^{\rho\alpha} +f^1_d q^\alpha v^\rho]
\gamma_5 \gamma^\mu_t
u_{\Lambda_{c1}} \; ,
\end{equation}
\begin{equation}
\label{coup-3}
M(\Lambda^*_{c1}\to\Sigma_c\gamma )=
\sqrt{3}e\epsilon_{\beta \nu \rho \mu} q^\beta {e^\nu}^*
{\bar u}_{\Sigma_c}
[f^2_s g_t^{\rho\alpha} +f^2_d q^\alpha v^\rho]
\gamma_5 \gamma_\alpha^t
u^\mu_{\Lambda_{c1}} \; ,
\end{equation}
\begin{equation}
\label{coup-4}
M(\Lambda^*_{c1}\to\Sigma^*_c\gamma )=
3e\epsilon_{\beta \nu \rho \mu} q^\beta {e^\nu}^*
{\bar u}^\alpha_{\Sigma^*_c}
[f^3_s g_t^{\rho\alpha} +f^3_d q^\alpha v^\rho]
u^\mu_{\Lambda_{c1}} \; ,
\end{equation}
where $e_\mu (\lambda )$ and $q_\mu$ are the photon polarization vector and momentum
respectively, $e$ is the charge unit. Due to heavy quark symmetry, we have
$f_s^1=f_s^2=f_s^3=f_s$, $f_d^1=f_d^2=f_d^3=f_d$. As in the case of pionic
couplings there are only two independent coupling constants associated with
the E1 and M2 decays.

We consider the correlator
\begin{equation}
\label{photon-1}
 \int d^4x\;e^{-ik\cdot x}\langle\gamma(q)|T\left(\eta_{\Lambda_{c1}}(0)
 {\bar \eta}_{\Sigma_c}(x)\right)|0\rangle =e
 {1+{\hat v}\over 2}  \gamma_\alpha^t \gamma^\mu_t
\epsilon_{\beta \nu \rho \mu} v^\beta {e^\nu}^*
\{F_s (\omega,\omega') g_t^{\rho\alpha} +
F_d (\omega,\omega') q^\alpha v^\rho \}
 \;,
\end{equation}
$F_{s,d} (\omega,\omega')$ has the same pole structures as $G_{s,d}(\omega,\omega')$.

The light cone two-particle photon wave functions are \cite{mendel}:
\begin{eqnarray}\label{wf-1}\nonumber
<\gamma (q)| {\bar q} (x) \sigma_{\mu\nu} q(0) |0>=
i e_q e \langle {\bar q} q\rangle
\int_0^1 du  e^{iuqx}\{ (e_\mu q_\nu -e_\nu q_\mu )  [\chi \phi (u) +x^2 h_1(u)]
&\\
+[ (qx)(e_\mu x_\nu -e_\nu x_\mu ) +(ex)(x_\mu q_\nu -x_\nu q_\mu )
-x^2(e_\mu q_\nu -e_\nu q_\mu )]h_2(u)\}  &\; ,
\end{eqnarray}
\begin{equation}\label{wf-2}
<\gamma (q)| {\bar q} (x) \gamma_\mu \gamma_5 q(0) |0>=
{f\over 4} e_q e \epsilon_{\mu\nu\rho\sigma}e^\nu q^\rho x^\sigma
\int_0^1 du  e^{iuqx} \psi (u) \; .
\end{equation}
The $\phi (u), \psi (u)$ is associated with the leading twist two photon
wave function, while $g_1(u)$ and $g_2(u)$ are twist-4 PWFs.
All these PWFs are normalized to unity, $\int_0^1 du \; f (u) =1$.

We want to emphasize that
the photon light cone wave functions include the complete perturbative and
non-perturbative electromagnetic interactions for the light quarks in principle.
Yet the interaction of the photon with the heavy quark is not parametrized
and constrained by the photon light cone wave functions. It seems possible that
the photon couples directly to the heavy quark line. This is different from the
QCD sum rules for the pionic couplings since pions can not couple directly
to the heavy quark. However the real photon coupling to heavy quark
involves a spin-flip transition, which is suppressed by a factor of
$1/m_Q$ \cite{korner}. So it vanishs in the leading order of $1/m_Q$ expansion.
Since we are interested in the leading order couplings $f_{s,d}$, it's
enough to keep the photon light cone wave functions for the light quarks only.

Expressing (\ref{photon-1}) with the photon wave functions, we arrive at:
\begin{eqnarray}\label{photon-2}\nonumber
F_s (\omega, \omega')=
{1\over \pi^2} (e_u -e_d)\langle {\bar q} q\rangle
\int_0^{\infty} dt \int_0^1 du
e^{i (1-u) \omega t } e^{i u \omega' t }
\{
 [{1\over  t^4} \chi \phi (u) &\\
+{1\over  t^2} (h_1(u) -h_2 (u))]
+{\pi^2\over 24}f \psi (u) t   \} +\cdots &  \;.
\end{eqnarray}
\begin{eqnarray}\label{photon-3}\nonumber
F_d (\omega, \omega')=
{i\over \pi^2} (e_u -e_d)\langle {\bar q} q\rangle
\int_0^{\infty} dt \int_0^1 du
e^{i (1-u) \omega t } e^{i u \omega' t } u
\{
 [{1\over  t^3} \chi \phi (u) &\\
+{1\over  t} (h_1(u) -h_2 (u))]
+{\pi^2\over 24} f\psi (u) t   \} +\cdots &  \;.
\end{eqnarray}

The final sum rules are:
\begin{eqnarray}\label{f-s}\nonumber
 f_s f_{\Lambda_{c1}} f_{\Sigma_c} =  -{a\over 4\pi^4} (e_u -e_d)
 e^{ {\bar\Lambda_{\Lambda_{c1}} +\bar\Lambda_{\Sigma_c}  \over 2T }}
 \{
\chi  \phi (u_0) T^5 f_4 ({s_0\over T})  &\\
-  [h_1(u_0)-h_2(u_0)] T^3f_2({s_0\over T})
+{\pi^2\over 24} f \psi (u_0)  T^1f_0 ({s_0\over T})
 \} &\;,
\end{eqnarray}
\begin{eqnarray}\label{f-d}
\nonumber
 f_d f_{\Lambda_{c1}} f_{\Sigma_c} =  -{a\over 4\pi^4} (e_u -e_d)
 e^{ {\bar\Lambda_{\Lambda_{c1}} +\bar\Lambda_{\Sigma_c}  \over 2T }} u_0
 \{
\chi \phi (u_0) T^4 f_3({s_0\over T}) &\\
- [h_1(u_0)-h_2(u_0)] T^2f_1({s_0\over T})
+{\pi^2\over 24} f \psi (u_0)  \} &\;.
\end{eqnarray}

\subsection{Numerical analysis of the photonic sum rules}

The leading photon wave functions receive only small corrections from
the higher conformal spins \cite{bely95} so they do not deviate
much from the asymptotic form. We shall use \cite{ali}
\begin{equation}
\phi (u) =6u(1-u) \; ,
\end{equation}
\begin{equation}
\psi (u) =1 \; ,
\end{equation}
\begin{equation}
h_1 (u) =-{1\over 8}(1-u) (3-u)\; ,
\end{equation}
\begin{equation}
h_2 (u) =-{1\over 4}(1-u)^2 \; .
\end{equation}
with $f=0.028$GeV$^2$ and $\chi =-4.4 $GeV$^2$ \cite{kogan,ioffe-mag,chiu,zhu-mag}
at the scale $\mu =1$GeV.

The variation of $f_{s,d}$ with the Borel parameter $T$ and $s_0$ is
presented in FIG. 3 and FIG. 4.
Stability develops for the sum rules (\ref{f-s}), (\ref{f-d})
in the region $0.5$ GeV $<$$T$$<$$1.5$ GeV, we get:
\begin{eqnarray}
\label{num-b-1}
 && f_s f_{\Lambda_{c1}} f_{\Sigma}=(2.0\pm 0.8)\times 10^{-4}\mbox{GeV}^6\;,\\
 &&f_d f_{\Lambda_{c1}} f_{\Sigma}=(4.8\pm 1.2)\times 10^{-4}\mbox{GeV}^5\;,
\end{eqnarray}
where the errors refers to the variations with $T$ and
$s_0$ in this region. And the central value corresponds
to $T=1.0$GeV and $s_0 =1.3$GeV. Our final result is
\begin{eqnarray}
\label{num-b-2}
 &&f_s  =(0.20\pm 0.08)\mbox{GeV}^{-1}\;,\\
 &&f_d  =(0.48\pm 0.12)\mbox{GeV}^{-2}\;.
\end{eqnarray}

The decay width formulas in the leading order of HQET are
\begin{eqnarray}
\label{widths}
&&\Gamma(\Lambda_{c1}\to \Sigma_c \gamma )=
16 \alpha |\vec q|^3 {m_{\Sigma_c}\over m_{\Lambda_{c1}}}
[f_s^2 +{1\over 2} f_d^2 |\vec q|^2 ]
 \;,\nonumber\\
&&\Gamma(\Lambda_{c1}\to \Sigma^*_c \gamma )=
8 \alpha |\vec q|^3 {m_{\Sigma^*_c}\over m_{\Lambda_{c1}}}
[f_s^2 +{1\over 2} f_d^2 |\vec q|^2 ]
 \;,\nonumber\\
&&\Gamma(\Lambda^*_{c1}\to \Sigma_c \gamma )=
4 \alpha |\vec q|^3 {m_{\Sigma_c}\over m_{\Lambda^*_{c1}}}
[f_s^2 +{1\over 2} f_d^2 |\vec q|^2 ]
 \;,\nonumber\\
&&\Gamma(\Lambda^*_{c1}\to \Sigma^*_c \gamma )=
20 \alpha |\vec q|^3 {m_{\Sigma^*_c}\over m_{\Lambda^*_{c1}}}
[f_s^2 +{1\over 2} f_d^2 |\vec q|^2 ]
 \;,
\end{eqnarray}
where $|\vec q|=134, 72, 164, 103$ MeV is the photon decay momentum
for the above four processes. The d-wave decay is negligible.
The decay width values are collected in TABLE III.
The uncertainty is typically about $50\%$.

The decays $\Lambda_{Q1}\to \Sigma_Q \gamma$ do not occur in the leading order
in the bound state picture \cite{chow}. Due to the unknown coupling constant
$c_{RS}$ in the chiral lagrangian for the heavy quark electromagnetic
interactions, no numerical values are available \cite{cho}. However the
decay width ratios of the four final states are exactly the same as
ours if we ignore the isospin violations of the heavy multiplet masses
in the heavy quark limit. Our results are much smaller than those from
various versions of quark models \cite{t-3,i-1}, which may indicate that
the $1/m_c$ correction is important.

\section{The process $\Lambda_{c1} \to \Lambda_c \gamma$ etc}
\label{sec5}

As can be seen later the radiative decay processes of p-wave $\Lambda_{c1}$
doublet to $\Lambda_c$ is quite different from those in the previous section.
We present more details here. The possible E1 decay amplitudes are
\begin{equation}
\label{c-1}
M(\Lambda_{c1}\to\Lambda_c\gamma )=
e h_p e^*_\mu {\bar u}_{\Lambda_c} [ g_t^{\mu\nu}v\cdot q - v^\mu q^\nu]
\gamma_\mu \gamma_5 u_{\Lambda_{c1}} \; ,
\end{equation}
\begin{equation}
\label{co-2}
M(\Lambda^*_{c1}\to\Lambda_c\gamma )=
\sqrt{3}e h'_p e^*_\mu {\bar u}_{\Lambda_c} [ g_t^{\mu\nu}v\cdot q - v^\mu q^\nu]
u_{\Lambda^\nu_{c1}} \; .
\end{equation}
Due to heavy quark symmetry $h_p=h'_p$.

We consider the correlator
\begin{equation}
\label{photon}
\Pi =i \int d^4x \;e^{ik\cdot x}\langle\gamma(q)|T\left(\eta_{\Lambda_{c1}}(x)
 {\bar \eta}_{\Lambda_c}(0)\right)|0\rangle = {1+{\hat v}\over 2}
 e^*_\mu [ g_t^{\mu\nu}v\cdot q - v^\mu q^\nu] \gamma_\mu \gamma_5
  H_p (\omega, \omega') \;.
\end{equation}

We first calculate the part solely involved with the light quark, which can be
expressed with the photon wave functions. We get
\begin{equation}
\label{photo}
\Pi = 2i \int_0^\infty dt \int d^4x\;e^{ik\cdot x} {\hat D}_t \delta (x-vt)
\gamma_5 {1+{\hat v}\over 2} \{ Tr [ \gamma_5 C iS_u^T (x) C\gamma_5
< \gamma (q) | d(x) \bar d (0) |0> ] + (u\leftrightarrow d ) \}  \; ,
\end{equation}
where summation over color has been performed.
There are two types of terms with even $\gamma$ matrices in the trace.
The first one is connected with $\psi (u)$ and the trace looks like
$ Tr [ \gamma_5 C {\hat x}^T  C \gamma_5 \gamma_\mu \gamma_5 ] $.
The second is involved with $\phi (u), h_1(u), h_2(u)$ and the trace
looks like  $Tr [ \gamma_5 C 1 C \gamma_5 \sigma_{\mu\nu} ]$.
In both $\Lambda_{Q1}, \Lambda_Q$ states the up and down quarks are
in the $0^+$ state, which leads to the presence of $\gamma_5 C$ and $C \gamma_5$
in both traces. Clearly both traces vanish.
This property results from the underlying flavor and spin structure of
the light quark sector. In other words the light quark contribution is zero
to all orders of the heavy quark expansion in the framework of LCQSR
with the commonly used interpolating currents (\ref{lab1}) and (\ref{la1})
for $\Lambda_Q$ and $\Lambda_{Q1}$ respectively.
The decays $\Lambda_{Q1} \to \Lambda_Q \gamma$ and
$\Lambda^*_{Q1} \to \Lambda_Q \gamma$ happens only when the photon
couples directly to the heavy quark line.

Now let's move to the part involved with the heavy quark.
At first sight there are two types of terms in the leading order of heavy
quark expansion. The first one comes from the insertion of the
operator $i\int \bar [h_v (y)iv\cdot D h_v (y)] d^4 y$ in (\ref{photon}), which contributes
a factor $v\cdot e^* (\lambda )$ to the decay amplitude.
For the real photon $v\cdot e^* (\lambda )=0$ so it drops out.
The other possible term arises from the covariant derivative in $\eta_{\Lambda_{c1}}$,
which leads to a nonzero correlator. For the tensor structure
$i{\hat e}^* \gamma_5 {1+{\hat v}\over 2}$ we have
\begin{equation}\label{qq}
\Pi (\omega, \omega') =-{e\over \pi^4}\int_0^\infty dt
e^{i\omega' t} \{
 { 6\over t^6} +{ <g_s^2 G^2 > \over 64 t^2} -{a^2\over 96} \} \;,
\end{equation}
where the photon field has contributed a factor $e^{-iq\cdot x}$.
It's important to note only the variable $\omega'$ appears in (\ref{qq}).
It's a single pole term which must vanish after we make double Borel transformation
to the variables $\omega, \omega'$ simultaneously. We have shown there is no
leading order E1 transition in (\ref{c-1}) arising from the photon
couplings to the heavy quark line in the leading order of heavy quark expansion.
Based on the same spin and flavor consideration we know that radiative
decay processes like $\Sigma_{Q1}\to \Lambda_{Q1} \gamma$,
$\Lambda_{Q1}\to \Lambda_{Q1} \gamma$, $\Sigma_{Q1}\to \Sigma_{Q1} \gamma$
are also forbidden in the leading order of $1/m_Q$ expansion, where we have used
notations in \cite{korner}.

We may rewrite the decay amplitudes as
\begin{equation}
\label{cc-1}
M(\Lambda_{c1}\to\Lambda_c\gamma )=
e f_p  F_{\mu\nu} {\bar u}_{\Lambda_c} \sigma^{\mu\nu} \gamma_5 u_{\Lambda_{c1}} \; ,
\end{equation}
\begin{equation}
\label{cco-2}
M(\Lambda^*_{c1}\to\Lambda_c\gamma )=
2\sqrt{3}e f^1_p F_{\mu\nu}v^\mu {\bar u}_{\Lambda_c} u_{\Lambda^\nu_{c1}} \; ,
\end{equation}
\begin{equation}
\label{cco-3}
M(\Lambda^*_{c1}\to\Lambda_{c1}\gamma )=
2\sqrt{3}e f^2_p F_{\mu\nu} {\bar u}_{\Lambda_{c1}} \gamma_t^\mu \gamma_5
u_{\Lambda^\nu_{c1}} \; .
\end{equation}
Due to heavy quark symmetry we have
\begin{equation}
f_p=f^1_p=f^2_p\; .
\end{equation}
Note $f_p ={1\over 4}h_p$.

In these decays we know the light quarks do not contribute. 
However, the $J^P$ of the light diquark changes from $1^-$ to $0^+$ which ensures
the decay $\Lambda_{c1}\to \Lambda_c \gamma$ is an E1 transition.
The angular momentum and parity $J^P={1\over 2}^+$ of the heavy quark does not change so 
the coupling constant $f_p$ is the same as that for the heavy quark $M1$ transition, which is
induced by the magnetic moment operator
\begin{equation}\label{fp}
f_p={\mu_c\over 2}={e_c\over 4m_c}\; .
\end{equation}

Another approach is to consider the three point correlation function for the 
tensor structure ${\hat e}_t \gamma_5 {1+{\hat v} \over 2}$
\begin{equation}\label{coo-1}
i\int d^4 x d^4 z e^{ikx-ik'z} \langle 0| T\{ \eta_{\Lambda_{c1}} (x),
{{\cal K}(0) +{\cal S}(0)\over 2m_c}, 
{\bar \eta}_{\Lambda_{c1}} (z)\} |0\rangle
=\Pi_3 (\omega, \omega') {1+{\hat v} \over 2}\;,
\end{equation}
with $\omega =k\cdot v, \omega'=k'\cdot v$.
\begin{equation}
\Pi_3 (\omega, \omega')={2e_c\over m_c}{1\over \pi^4}\int_0^\infty dt_1 dt_2 
e^{i\omega t_1+i\omega' t_2} \{ {18\over (t_1+t_2)^8} +
{<g_s^2 G^2>\over 64 (t_1+t_2)^4} \} \; ,
\end{equation}
After the double Borel transformation and continuum subtraction we get the 
sum rule for $h_p$
\begin{equation}\label{hp}
h_p (\bar \Lambda_{\Lambda_{c1}} -\bar \Lambda_{\Lambda_c}) f_{\Lambda_{c1}}f_{\Lambda_c}
e^{-{\bar \Lambda_{\Lambda_{c1}} +\bar \Lambda_{\Lambda_c}\over 2T}} =
{1\over \pi^4}{e_c\over m_c} \{ 36 T^8 f_7 ({s_0\over T}) +{<g_s^2 G^2>\over 32}
T^4 f_3 ({s_0\over T}) \} \; .
\end{equation}
Dividing (\ref{hp}) by (\ref{mass-1}) we get
\begin{equation}\label{hphp}
h_p={e_c\over m_c} e^{-{\bar \Lambda_{\Lambda_{c1}} -\bar \Lambda_{\Lambda_c}\over 2T}} 
{f_{\Lambda_{c1}}\over 3(\bar \Lambda_{\Lambda_{c1}} -\bar \Lambda_{\Lambda_c}) f_{\Lambda_c}}
{ T^8 f_7 ({s_0\over T}) +{<g_s^2 G^2>\over 1152}T^4 f_3 ({s_0\over T}) \over
T^8 f_7 ({s_0\over T}) -{<g_s^2 G^2>\over 6912}T^4 f_3 ({s_0\over T})
+{m_0^2 a^2 \over 8192} } \; .
\end{equation}
Numerically we have $h_p \approx {e_c\over m_c}$, which is consistent with (\ref{fp}).
The decay widths formulas are
\begin{eqnarray}
\label{widths-p}
&&\Gamma(\Lambda_{c1}\to \Lambda_c \gamma )=
e_c^2 \alpha  |\vec q|^3 {m_{\Lambda_c}\over m_{\Lambda_{c1}}m_c^2}
 \;,\nonumber\\
&&\Gamma(\Lambda^*_{c1}\to \Lambda_c \gamma )=
e_c^2 \alpha  |\vec q|^3 {m_{\Lambda_c}\over m_{\Lambda^*_{c1}}m_c^2}
 \;,\nonumber\\
&&\Gamma(\Lambda^*_{c1}\to \Lambda_{c1} \gamma )=
e_c^2 \alpha  |\vec q|^3 {m_{\Lambda_{c1}}\over m_{\Lambda^*_{c1}}m_c^2}
 \;.
\end{eqnarray}
The decay momentum is $290, 320, 32$ MeV respectively. We take $m_c=1.4$ GeV.
The numerical values are collected in TABLE III.
These widths comes solely from the ${\cal O} (1/m_Q)$ correction.
But their numerical values are greater than those leading order widths for
the channels $\Sigma_c\gamma, \Sigma^*_c\gamma$. The reason is purely
kinematical. The decay momentum for the final state $\Lambda_c\gamma$
is three times larger. For the p-wave decay there appears an enhancement
factor of 27.

These widths in (\ref{widths-p}) are propotional to ${e_c^2\over m_c^2}$. Therefore the
corresponding radiative decays $\Lambda_{b1}\to \Lambda_b\gamma$,
$\Lambda^*_{b1}\to \Lambda_b\gamma$, $\Lambda^*_{b1}\to \Lambda_{b1}\gamma$
are further suppressed by a factor $({e_b\over e_c} {m_c\over m_b})^2 \sim 40 $.
The widths of the first two decays are around $1$ keV.

If we use naive dimensional analysis to let $c_{RT}$ \cite{cho} in TABLE III
be of the order of unity or simply assume that the E1 transition coupling constant
$h_p$ in (\ref{c-1}) is of the same order of M1 transition one \cite{korner},
we would get a width ${\cal O}(100)$ keV. Our result is in strong contrast
with those from the bound state picture \cite{chow}, where $\Gamma (\Lambda_{c1},
\Lambda^*_{c1}\to \Lambda_c \gamma)=16, 21$ keV and $\Gamma (\Lambda_{b1},
\Lambda^*_{b1}\to \Lambda_b \gamma)=90, 119$ keV. Future experiments
should be able to judge which mechanism is correct.

It was noted that the radiative decays $\Lambda_{Q1} \to \Lambda_Q$ was forbidden
in the leading order of heavy quark symmetry assuming one-body transition operators,
which arises from a complete cancellation due to the specific spins of light
constituent quarks in the antisymmetric initial and final state \cite{t-3}.
The point is consistent with our observation of the vanishing contribution
of the light quark sector to this radiative process.

>From our calculation we know
the d-wave single pion width of $\Lambda^*_{c1}$ is $33$ keV and the estimate
in \cite{dai} yielded $35$ keV for the two pion decay width. It's
interesting to notice that the radiative decay widths are $48, 5, 6, 0.014$ keV
for the final states $\Lambda_c \gamma, \Sigma_c\gamma, \Sigma_c^*\gamma, \Lambda_{c1}\gamma$
respectively. The width of the decay channel $\Lambda_c \gamma$
is bigger than either of that of the strong decay modes.
The $\Lambda_{c1}^*$ should be a narrow state with a total width about $130$ keV.

The two pion width of $\Lambda_{c1}$ is about $2.5$ MeV \cite{dai}.
>From TABLE II and III the one pion and electromagnetic widths are
$\Gamma(\Sigma_c\pi, \Lambda_c \gamma, \Sigma_c\gamma, \Sigma^*\gamma)
=2.7, 0.048, 0.011, 0.001$ MeV. Its total width is about $5.4$ MeV.

It's believed that $\Lambda_{b1}$ lies below $\Sigma_b \pi, \Lambda_b \pi\pi$ threshold.
If so its dominant decays are electromagnetic. From our calculation we see
$\Gamma (\Lambda_{b1}\to \Lambda_b \gamma, \Sigma_b\gamma, \Sigma_b^*\gamma)
=1, 11, 1$ keV if we assume the same decay momentum as in the $\Lambda_{c1}$
decays. Its total width is about $13$ keV. It will be a very narrow state.
Clearly the radiative channels $\Sigma_b\gamma$
will be very useful to find them experimentally.

The major decay modes of $\Lambda^*_{b1}$ might be d-wave one pion decay and
electromagnetic decays to $\Sigma_b$ doublet if the two pion mode is not
allowed. Their widths are $\Gamma (\Lambda^*_{b1}\to\Sigma_b\pi,
\Lambda_b \gamma, \Sigma_b\gamma, \Sigma_b^*\gamma)
=33, 1, 5, 6$ keV if we assume the same decay momentum as in the
$\Lambda_{c1}^*$ case. It's also a very narrow state with a width of $45$ keV.

Before ending this section we want to improve our previous calculation of
radiative decays of excited heavy mesons \cite{zhu-rad}. (1) First the s-wave
terms involved with $g_s$ should not appear in $(1^+, 2^+)\to (0^-, 1^-) \gamma$ processes.
All decays are M2 transitions. The $g^2_s$ in the decay width formulas should be
replaced by ${1\over 9}g_d^2 |{\vec q}|^4$. The last eight widths in Eq. (94)
should read $2, 8, 3, 11, 6, 23, 7, 27$ keV respectively, which
is much smaller than original wrong ones.
(2) The E1 transition $(0^+, 1^+)\to (0^-, 1^-) \gamma$
decays was identified as s-wave decays. This was misleading.
The factor $(q\cdot v)$ should be in the tensor structure to ensure the
E1 transition structure in Eq. (47) in \cite{zhu-rad}. We present
the correct sum rules for $g_1$ below.
\begin{equation}
g_1 f_{-,1/2}f_{+,1/2} =-{a\over 4\pi^2} e^{\Lambda_{-,1/2}+\Lambda_{+,1/2}\over 2T}
\{ \chi \phi (u_0) Tf_0 ({s_0\over T}) -g_1(u_0) {1\over T} \} \;,
\end{equation}
where $s_0=\omega_c /2 =(1.5\pm 0.2) $ GeV. Numerically we have $g_1 =(1.6\pm 0.2)$
GeV$^{-1}$.

\section{Discussions}
\label{sec6}

In our calculation only the errors due to the variations of $T$ and $s_0$
are included in the final results for $g_{s,d}$, $f_{s,d}$.
The various input parameters like quark condensate, gluon condensate,
$\chi$, $f$ etc also have some uncertainty. Among these the values of
the pion and photon wave functions introduce largest uncertainty. Although their values are
constrained by either experimental data or other QCD sum rule analysis, 
they may still lead to $\sim 25\%$ unceritainty.
Keeping the light cone wave functions up to twist four also leads to some errors.
However the light cone sum rules are dominated by the lowest twist wave functions.
Take the sum rule (\ref{f-d}) for $f_d$ for an example.
At $T=1$ GeV, the twist-four term involved with $h_1, h_2$ is only
$9\%$ of the leading twist term after the continuum subtraction.
In other words the light cone expansion converges quickly. So
we expect the contribution of higher twist terms to be small.
There are other sources of uncertainty which is difficult to estimate.
One is the QCD radiative correction, which is not small in
both the mass sum rule and LCQSRs for the pionic coupling constants
of the ground state heavy hadrons in HQET. But their ratio depends
only weakly on these corrections because of large cancellation \cite{report}.
Numerically the radiative corrections are around $10\%$ of the tree 
level result.  

Another possible source is the $1/m_Q$ correction for the charmed p-wave
baryons. The leading order coupling constants $g_{s,d}$ etc will be corrected
by terms like $g'_{s,d}/m_Q$, which will affect decay widths.
For the charmed hadrons $1/m_Q$ corrections
are sizable and may reach $30\%$ while such corrections are generally
less than $10\%$ of the leading order term for the bottom system \cite{zhu-meson}.
Especially for the E1 transition coupling constant $f_s$, the correction is of the
order ${e_c\over 4m_c}$, which may be comparable with the leading order one for
the charm system. One is justified to use these coupling constants to calculate
the decay widths of the p-wave bottom baryons.
Unfortunately data is still not available for the p-wave bottom baryons.
So we have calculated the p-wave $\Lambda_{c1}$ doublet decay widths with
some reservation.

In summary we have calculated the pionic and electromagnetic coupling
constants and decay widths of the lowest p-wave heavy baryon doublet.
We compare our calculation with different approaches in literature.
We hope these results will be useful in the future experimental search of
$\Lambda_{b1}, \Lambda_{b1}^*$ baryons.


\vspace{0.8cm} {\it Acknowledgements:\/} S.-L.Z. is grateful to Prof.
C.-S. Huang for bringing the topic of excited heavy baryon to his attention.
\bigskip
\vspace{1.cm}


\newpage

\begin{table}[t]
\caption{Pionic coupling constants}
\begin{tabular}{|c|c|c|c|}
Coupling & Our & Ref. \cite{i-1} & Ref. \cite{t-2}\\
\hline\hline
$g_s$ &$0.5\pm 0.3$ & 0.52 & 0.665$\pm$0.135\\
$3g_d$ &$(8.4\pm 1.8)$GeV$^{-2}$& 21.5 GeV$^{-2}$&50.85$\pm$14.25 GeV$^{-2}$\\
\end{tabular}
\end{table}

\begin{table}[t]
\caption{Single pion decay widths}
 \begin{tabular}{|c|c|c|c|c|c|}
 & Our& Ref. \cite{dai}& Ref.\cite{i-1} & Ref. \cite{t-2}  &  Experiment  \\
\hline \hline
\multicolumn{4}{|l|}{S-wave transitions} \\
\hline
$\Lambda_{c1; S}(2593) \rightarrow \Sigma^{0}_c\pi^{+} $ &0.86 MeV & $0.89\pm 0.86$ MeV & $0.83\pm 0.09  $ MeV
& $ 1.775\pm 0.695$ MeV
& $0.86^{+0.73}_{-0.56}$ MeV\\
$\Lambda_{c1; S}(2593) \rightarrow \Sigma^{+}_c\pi^{0} $ & 1.2 MeV&$1.7\pm 0.49$ MeV &$0.98\pm 0.12  $ MeV
& $1.18\pm 0.46$ MeV
& $ \Gamma({\Lambda_{c1; S}}) = 3.6^{+2.0}_{-1.3}$ MeV \\
$\Lambda_{c1; S}(2593) \rightarrow \Sigma^{++}_c\pi^{-} $ &0.64 MeV &$0.55\pm ^{1.3}_{0.55}$ MeV  & $0.79\pm 0.09$ MeV&
$ 1.47\pm 0.57$ MeV
& $0.86^{+0.73}_{-0.56}$ MeV\\
\hline \hline
\multicolumn{4}{|l|}{D-wave transitions} \\
\hline
$\Lambda^{*}_{c1; S}(2625) \rightarrow \Sigma^{0}_c\pi^{+} $
& 0.011 MeV& 0.013 MeV& $ 0.080\pm 0.009 $ MeV
& $0.465\pm 0.245$ MeV& $< 0.13$ MeV\\
$\Lambda^{*}_{c1; S}(2625) \rightarrow \Sigma^{+}_c\pi^{0} $
&0.011 MeV&0.013 MeV& $ 0.095\pm 0.012 $ MeV
& $ 0.42\pm 0.22$ MeV& $\Gamma({\Lambda^{*}_{c1}}) <1.9 $MeV  \\
$\Lambda^{*}_{c1; S}(2625) \rightarrow \Sigma^{++}_c\pi^{-} $
& 0.011MeV&0.013 MeV& $ 0.076\pm 0.009 $ MeV
& $ 0.44\pm 0.23$ MeV& $<0.15$ MeV\\
\end{tabular}
\end{table}

\begin{table}[t]
\caption{Radiative decay widths}
\begin{tabular}{|c|c|c|c|c|c|}
 &Our&Ref. \cite{t-3} & Ref. \cite{i-1} & Others  & Experiment \cite{pdg}\\
\hline
$\Lambda_{c1; S}(2593) \rightarrow \Lambda_c^+\gamma $ & 0.036 MeV&0& $0.115\pm 0.001$ MeV
& $0.191c^2_{RT}$ MeV \cite{cho} & $< 2.36^{+1.31}_{-0.85}$ MeV \\
& & & $0.016$ MeV \cite{chow} &\\
\hline
$\Lambda_{c1; S}(2593) \rightarrow \Sigma_c^+\gamma $ & 0.011 MeV&0.071 MeV& $0.077\pm 0.001$ MeV
& $0.127c^2_{RS}$ \cite{cho} & \\
\hline
$\Lambda_{c1; S}(2593) \rightarrow \Sigma_c^{*+}\gamma $&0.001 MeV &0.011 MeV& $0.006\pm 0.0001$
MeV & $0.006c^2_{RS}$ \cite{cho} & \\
\hline
$\Lambda_{c1; S}^*(2625) \rightarrow \Lambda_c^+\gamma$&0.048 MeV&0 &$0.151\pm 0.002$ MeV
& $0.253c^2_{RT}$ MeV \cite{cho} & $< 1$ MeV\\
 && & $0.021$ MeV \cite{chow} &\\
\hline
$\Lambda_{c1; S}^*(2625) \rightarrow \Sigma_c^+\gamma $&0.005 MeV&0.13 MeV & $0.035\pm 0.0005$ MeV
& $0.058c^2_{RS}$ \cite{cho} & \\
\hline
$\Lambda_{c1; S}^*(2625) \rightarrow \Sigma_c^{*+}\gamma $ &0.006 MeV&0.032 MeV
& $0.046\pm 0.0006$ MeV & $0.054c^2_{RS}$ \cite{cho} & \\
\end{tabular}
\end{table}

\newpage
{\bf Figure Captions}
\vspace{2ex}
\begin{center}
\begin{minipage}{120mm}
{\sf
Fig. 1.} \small{Dependence of $g_s f_{\Lambda_{c1}} f_{\Sigma_c}$ on the Borel parameter $T$ for
different values of the continuum threshold $s_0$.
>From top to bottom the curves correspond
to $s_0=1.4, 1.3, 1.2$ GeV. }
\end{minipage}
\end{center}
\begin{center}
\begin{minipage}{120mm}
{\sf
Fig. 2.} \small{Dependence of $g_d f_{\Lambda^*_1} f_{\Sigma_c}$ on $T$, $s_0$. }
\end{minipage}
\end{center}
\begin{center}
\begin{minipage}{120mm}
{\sf
Fig. 3.} \small{Dependence of $f_s f_{\Lambda_{c1}} f_{\Sigma_c}$ on $T$, $s_0$. }
\end{minipage}
\end{center}
\begin{center}
\begin{minipage}{120mm}
{\sf
Fig. 4.} \small{Dependence of $f_d f_{\Lambda_{c1}} f_{\Sigma_c}$ on $T$, $s_0$. }
\end{minipage}
\end{center}

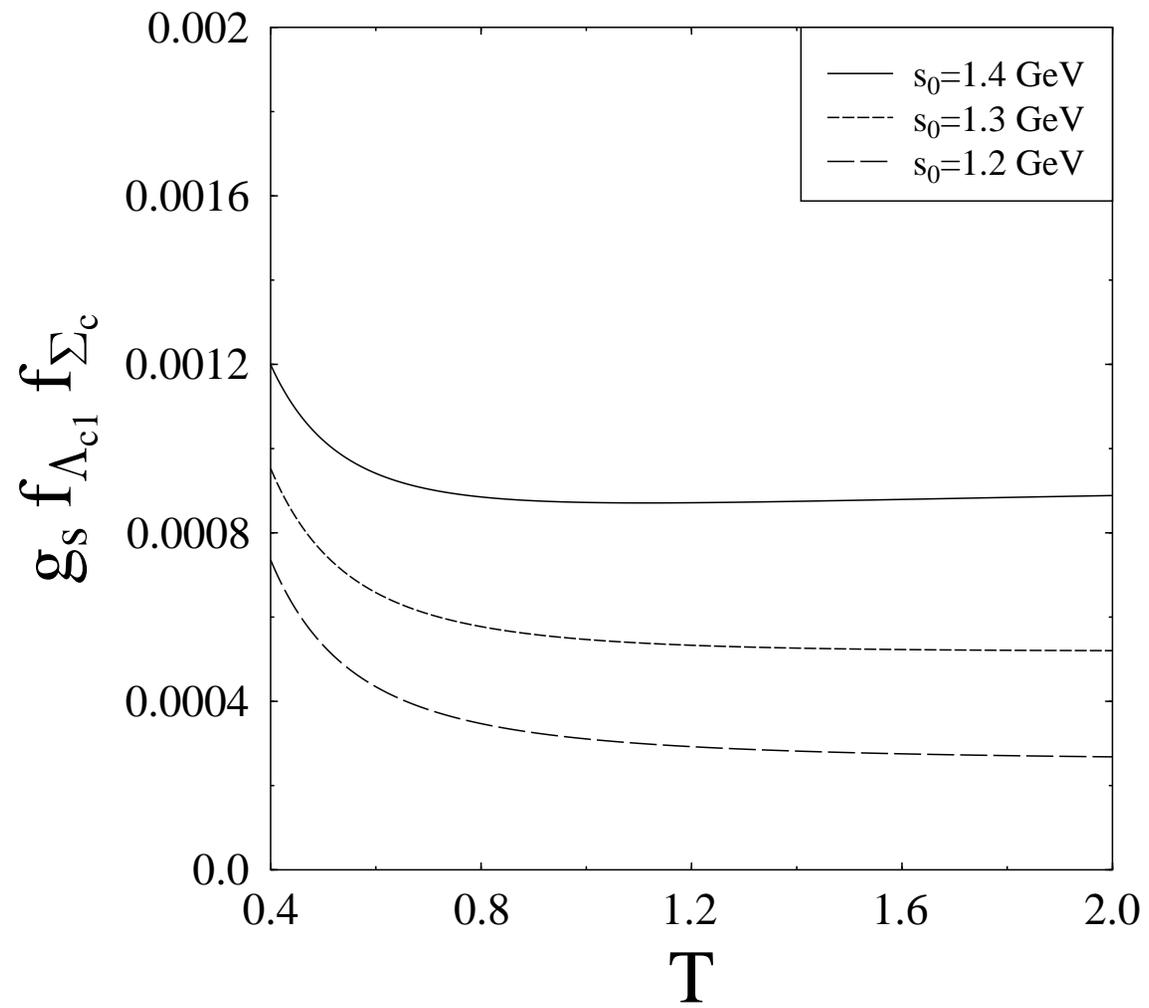
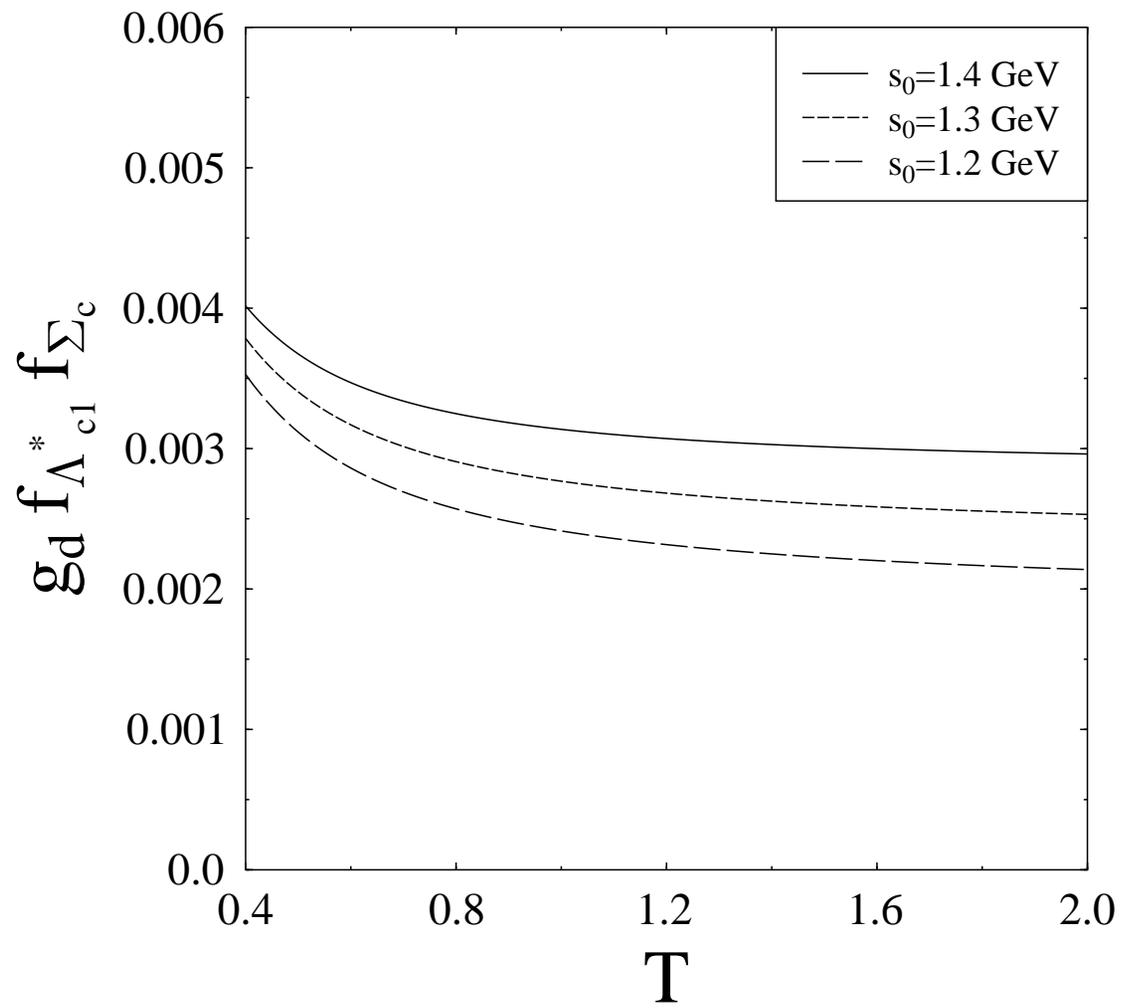
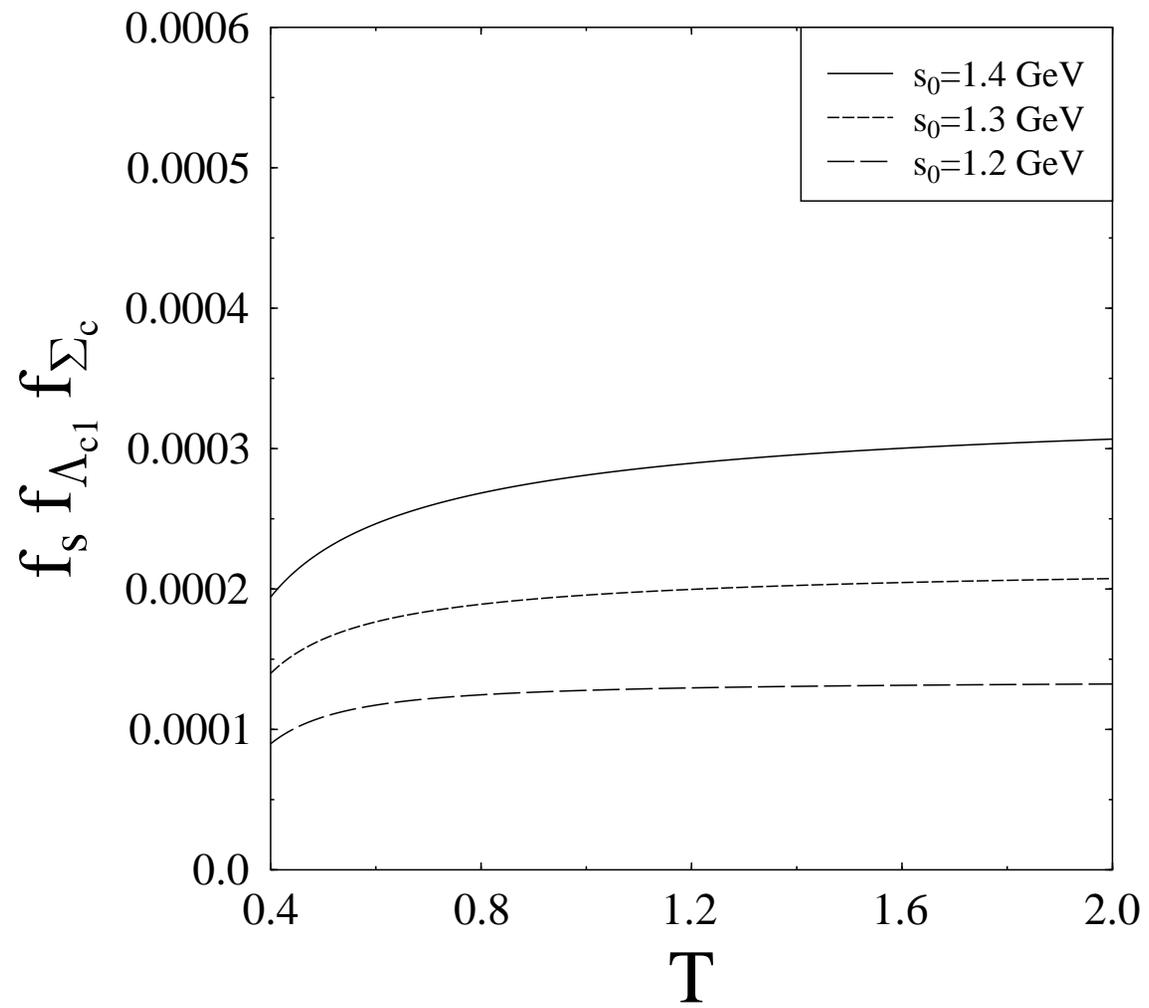
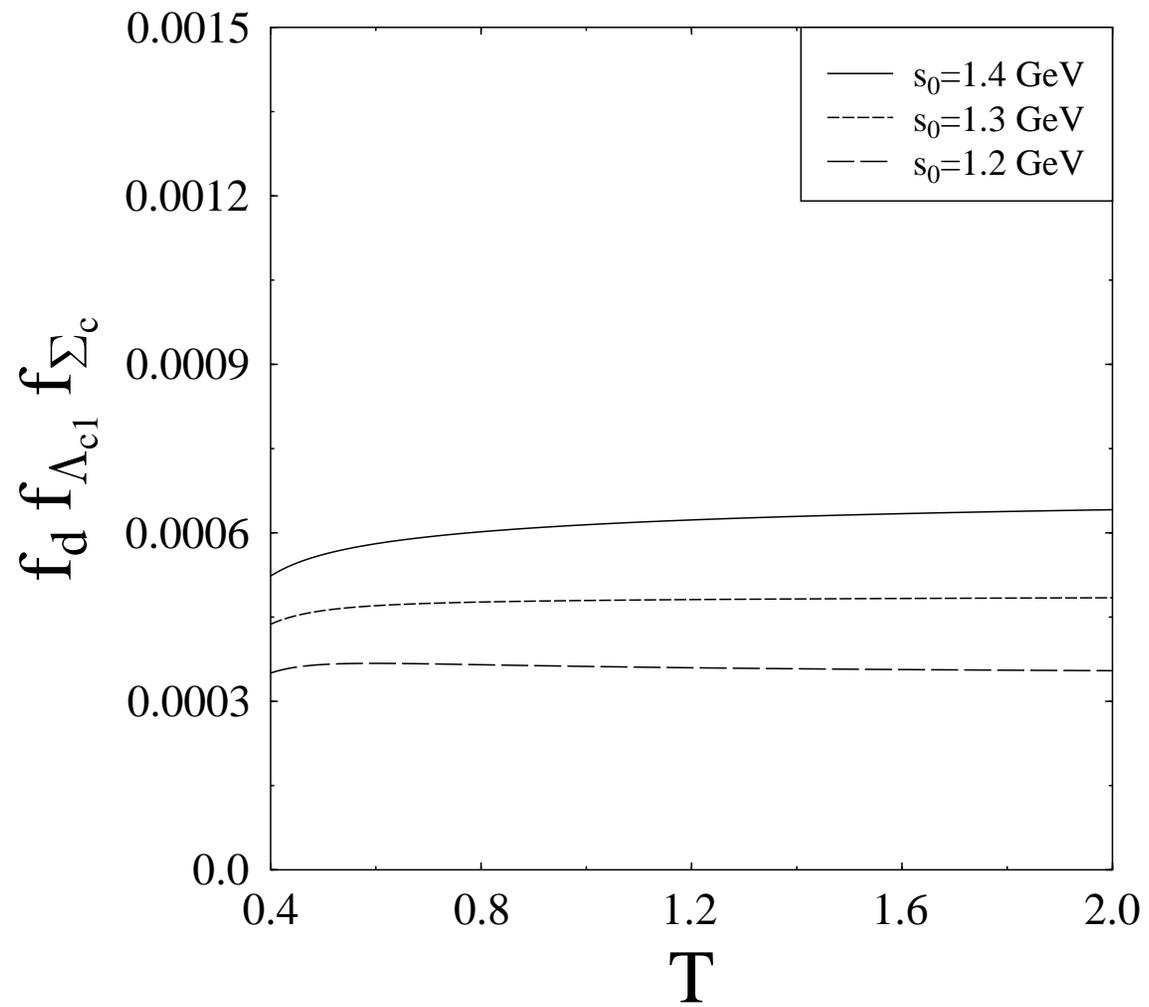
\end{document}